\RequirePackage{arydshln}
\documentclass[aps,twocolumn,nofootinbib,superscriptaddress,preprintnumbers,pra,10pt,floatfix]{revtex4-1}

\usepackage[utf8]{inputenc}

\usepackage{float}
\usepackage{amsmath,amssymb}
\usepackage{dsfont} 
\usepackage{hyperref}
\usepackage{graphicx}
\usepackage{enumitem}
\usepackage{mathtools}
\usepackage{bbold}
\usepackage{multirow}
\usepackage{ytableau}
\usepackage{youngtab}
\usepackage{braket}
\usepackage{soul}
\usepackage{cancel}
\usepackage{xcolor}
\usepackage[normalem]{ulem}
\usepackage{hhline}

\usepackage{lipsum}

\usepackage{colortbl} 
\definecolor{Gray}{gray}{0.95}
\definecolor{RGray}{gray}{0.90}
\definecolor{CGray}{gray}{0.92}
\definecolor{ForestGreen}{RGB}{34, 139, 34} 

\usepackage{arydshln}

\usepackage{tikz}
\usetikzlibrary{calc,tikzmark,fit,shapes.geometric,matrix,decorations.markings,arrows.meta,decorations.pathmorphing,patterns,positioning,snakes}

\tikzset
  {midarrow/.style={decoration={markings,mark=at position 0.5 with
     {\arrow[thin,xshift=2pt]{Triangle[length=4pt,#1]}}},postaction={decorate}}
  }

\tikzset{
proton/.style = {circle, draw=black, thin, fill=black!20!white, minimum size=#1,
              inner sep=0pt, outer sep=0pt},
proton/.default = 6pt 
}

\tikzset{
blob/.style = {circle, draw=black, thin, preaction={fill, black!20!white}, pattern=north east lines, minimum size=#1,
              inner sep=0pt, outer sep=0pt},
blob/.default = 6pt 
}

\tikzset{
wc/.style = {circle, fill, minimum size=#1,
              inner sep=0pt, outer sep=0pt},
wc/.default = 4pt 
}

\tikzset{vector/.style={decorate, decoration=snake}}

\makeatletter
\g@addto@macro\bfseries{\boldmath}
\makeatother

\makeatletter
\renewcommand\paragraph{\@startsection{paragraph}{4}{\z@}%
                                    {3.25ex \@plus1ex \@minus.2ex}%
                                    {-1em}%
                                    {\normalfont\normalsize\bfseries}}
\makeatother

\allowdisplaybreaks
\newcommand{\cB}{\mathcal{B}}
\newcommand{\cA}{\mathcal{A}}
\newcommand{\cL}{\mathcal{L}}

\begin{document}

\preprint{CERN-TH-2024-183}
\preprint{ZU-TH 53/24}

\title{ 
Probing third-generation New Physics 
with $K\to \pi \nu\bar\nu$ and 
$B\to K^{(*)} \nu\bar\nu$}

\author{L.~Allwicher}
\email{lukall@physik.uzh.ch}
\affiliation{Physik-Institut, Universit\"{a}t Z\"{u}rich, CH-8057 Z\"{u}rich, Switzerland}
\author{M.~Bordone}
\email{marzia.bordone@cern.ch}
\affiliation{Theoretical Physics Department, CERN, Geneva, Switzerland}
\author{G.~Isidori}
\email{gino.isidori@uzh.ch}
\affiliation{Physik-Institut, Universit\"{a}t Z\"{u}rich, CH-8057 Z\"{u}rich, Switzerland}
\author{G.~Piazza}
\email{gioacchino.piazza@physik.uzh.ch}
\affiliation{Physik-Institut, Universit\"{a}t Z\"{u}rich, CH-8057 Z\"{u}rich, Switzerland}

\author{A.~Stanzione}
\email{alfredo.stanzione@sissa.it}
\affiliation{
INFN, Sezione di Trieste, SISSA, Via Bonomea 265, 34136, Trieste, Italy}
\affiliation{SISSA International School for Advanced Studies, Via Bonomea 265, 34136, Trieste, Italy}

\begin{abstract}
\vspace{5mm}
The recent observation of the $K^+ \to \pi^+ \nu\bar\nu$ decay by NA62 is an important milestone in precision flavor physics. Together with evidence of $B^+ \to K^+\nu\bar\nu$ reported by Belle-II, they are the only FCNC decays involving third-family leptons where a precision close to the SM expectation has been reached. We study the implications of these recent results in the context of a new physics scenario aligned to the third generation, with an approximate $U(2)^5$ flavor symmetry acting on the light families. We find that the slight excess observed in both channels supports the hypothesis of non-standard TeV dynamics of this type, as also hinted at by other $B$-meson decays, consistently with bounds from colliders and electroweak observables. We further discuss how future improvements in precision could affect this picture, highlighting the discovery potential in these di-neutrino modes.

\vspace{3mm}
\end{abstract}

\maketitle

\allowdisplaybreaks

\section{Introduction}\label{sec:intro}

Direct and indirect searches for physics beyond the Standard Model (SM) impose stringent constraints on models addressing the electroweak hierarchy problem, which call  for new degrees of freedom at TeV energies. As shown in Ref.~\cite{Allwicher:2023shc}, a consistent New Physics (NP) framework compatible with current  data arises if we assume that the new dynamics predominantly couples to third-generation fermions while exhibiting weaker and flavor-universal interactions with the lighter families. This general framework is motivated both by the observed mass hierarchies and the electroweak hierarchy problem, as well as  by the lack of significant deviations in flavor-changing processes along with the relative weakness of bounds from direct searches that focus solely on third-generation fermions.

The analysis of Ref.~\cite{Allwicher:2023shc} has highlighted the interplay of constraints from direct searches, electroweak precision observables (EWPO), and flavor-changing processes in constraining and possibly detecting this framework. Among flavor observables, a very interesting role is played by  $K\to \pi \nu\bar\nu$ and $B\to K^{(*)} \nu\bar\nu$.
 These are the only flavor-changing neutral current (FCNC) transitions involving third-generation lepton pairs (specifically, tau neutrinos) for which current measurements have achieved the necessary sensitivity to probe the corresponding SM predictions. 
Motivated by recent experimental results on these modes~\cite{NA62talk,Belle-II:2023esi}, particularly by the clear observation of 
$K^+\to \pi^+ \nu\bar\nu$ by NA62~\cite{NA62talk}, in this paper we analyze the present and near-future impact of these modes in constraining such motivated class of NP models. 

We analyze the problem employing a general Effective Field Theory approach (EFT), focusing on dimension-six semileptonic operators built in terms of SM fields. The class of models we are interested in is characterized by a minimally-broken $U(2)^5$ flavor symmetry acting on the lightest two SM families~\cite{Barbieri:2011ci,Barbieri:2012uh,Isidori:2012ts},
which allows us to 
reduce the number of  relevant operators (and corresponding free parameters).
As we shall show, using only five parameters we can analyze consistently a large set of observables that includes direct searches, EWPO, and rare flavor-changing processes.
Within this context, we illustrate the key role of the two neutrino modes in determining the allowed parameter space.  
This exercise is particularly interesting since all the 
flavor observables, namely the lepton universality ratios
$R_{D}$ and $R_{D^{*}}$, the FCNC decays   
$B \to K^{(*)} \mu\bar \mu$, and the two neutrino modes, exhibit deviations from the corresponding SM predictions. While none of them is statistically very significant at present, their combination provides and interesting coherent picture.   

The paper is organized as follows.
Before discussing possible NP effects, in Section~\ref{sect:SM} we briefly review the SM predictions for $\cB(K^+\to\pi^+\nu\bar\nu)$ and $\cB(K_L \to\pi^0\nu\bar\nu)$, focusing in particular on the uncertainty related to $|V_{cb}|$.
In Section~\ref{sect:EFT} we introduce the EFT framework and in Section~\ref{sect:numerics} we proceed with the numerical analysis. 
There we also compare our findings with explicit NP models and discuss future prospects. 
The results are summarized in the Conclusions.
Numerical values and analytical expressions for all the 
observables considered in Section~\ref{sect:numerics} are reported in Appendix~\ref{sect:appendix}.

\section{SM predictions for $\cB(K\to\pi\nu\bar\nu)$}
\label{sect:SM}
Within the SM, short-distance contributions to the
$K\to \pi\nu\bar\nu$ and 
$B\to K^{(*)}\nu\bar\nu$ decays are described by the following effective Lagrangian
\begin{align}\label{eq:Leff}
\cL_\mathrm{eff} = \frac{4 G_\mathrm{F}}{\sqrt{2}} \frac{\alpha} {2 \pi} 
 \sum_{\ell = e,\mu,\tau}
\Big[ & \lambda^t_{sd} C^{\rm SM}_{\ell,sd}~ O^{\nu}_{\ell,sd}\nonumber\\ 
 + & \lambda^t_{bs} C^{\rm SM}_{\ell,bs}~ O^{\nu}_{\ell,bs} \Big]  +
\mathrm{h.c.}\,,
\end{align}
where 
\begin{equation}
    \begin{aligned}
\lambda^{q}_{ij} &=V^*_{qi}V_{qj}\,,
\qquad     
& O^{\nu}_{\ell,ij} = 
(\bar{d^i}_L\gamma_\mu d^j_L)(\bar{\nu}^\ell_L\gamma^\mu \nu^\ell_L)\,,
\nonumber\\
C^{\rm SM}_{\ell,bs} &= - \frac{1}{s_W^2} X_t\,, 
&C^{\rm SM}_{\ell,sd} = 
+ C^{\rm SM}_{\ell,bs} - \frac{ \lambda^c_{sd} }{s_W^2 \lambda^t_{sd} } X^\ell_c\,, 
    \end{aligned}
\end{equation}
$V_{ij}$ denotes CKM matrix elements, and the loop functions 
$X_t$ and $X^\ell_c$ have been computed  in~\cite{Buchalla:1998ba,Brod:2008ss,Buras:2006gb,Brod:2010hi}.
The leading top-quark contribution yields
\begin{equation}
    X_t = 1.48 \pm 0.01\,,\qquad
    C^{\rm SM}_{\ell,bs} = - 6.32 \pm 0.07\,.
\end{equation}

In the case of $K^+\to \pi^+\nu\bar\nu$ and $K_L \to \pi^0 \nu\bar\nu$,
extracting the hadronic matrix element from leading semileptonic $K$ decays, and 
taking into account various subleading corrections, leads to the following phenomenological expression~\cite{Brod:2021hsj}
\begin{align}\label{eq:Ktopi}
    &\mathcal{B}(K^+\to \pi^+\nu\bar\nu)= \,\kappa_+ (1+\Delta_\mathrm{EM})  \times \nonumber\\ 
    &\bigg[\bigg(\frac{\mathrm{Im}\lambda^t_{sd}}{\lambda^5} X_t\bigg)^2+\bigg(\frac{\mathrm{Re}\lambda^c_{sd}}{\lambda} (P_c+\delta P_{c,u})+\frac{\mathrm{Re}\lambda^t_{sd}}{\lambda^5} X_t\bigg)^2\bigg]\,, \nonumber\\
    &\mathcal{B}(K_L\to \pi^0\nu\bar\nu)= \, \kappa_L r_{\epsilon_K} \bigg(\frac{\mathrm{Im}\lambda^t_{sd}}{\lambda^5}X_t\bigg)^2\,.
\end{align}
where $\lambda \doteq |V_{us}|$.
Here $\delta P_{c,u}$ denotes the corrections from dimension-eight operators and long-distance contributions~\cite{Isidori:2005xm,Lunghi:2024sjy}, 
$\Delta_\mathrm{EM}$ encodes the effect of NLO QED corrections~\cite{Mescia:2007kn}, $r_{\epsilon_K}$ takes into account the indirect CP-violating contribution~\cite{Brod:2021hsj}, 
while the impact of the hadronic matrix element is encoded in  the  normalization factors $\kappa_{+,L}$. 
The numerical value of these coefficients are reported in Appendix~\ref{sect:appendix}.

\subsection{The impact of $|V_{cb}|$}
At present, the leading source of uncertainty in predicting $K\to \pi\nu\bar\nu$ rates within the SM lies in the CKM inputs and, particularly, on $|V_{cb}|$.
The dependence from $|V_{cb}|$ is hidden in the factor $\lambda^t_{sd}$ that, employing the improved Wolfenstein parametrization~\cite{Buras:1994ec},
can be rewritten as 
\begin{equation}
\lambda^t_{sd} =   \lambda |V_{cb}|^2   \left[ (\bar{\rho}-1) \left(1-\frac{\lambda^2}{2}\right) 
+ i \bar\eta \left(1+\frac{\lambda^2}{2}\right)  \right]\,,
\end{equation}
implying $\cB(K\to \pi\nu\bar\nu) \propto |V_{cb}|^4$.

\begin{figure}
\centering
\includegraphics[width = 8.5cm]{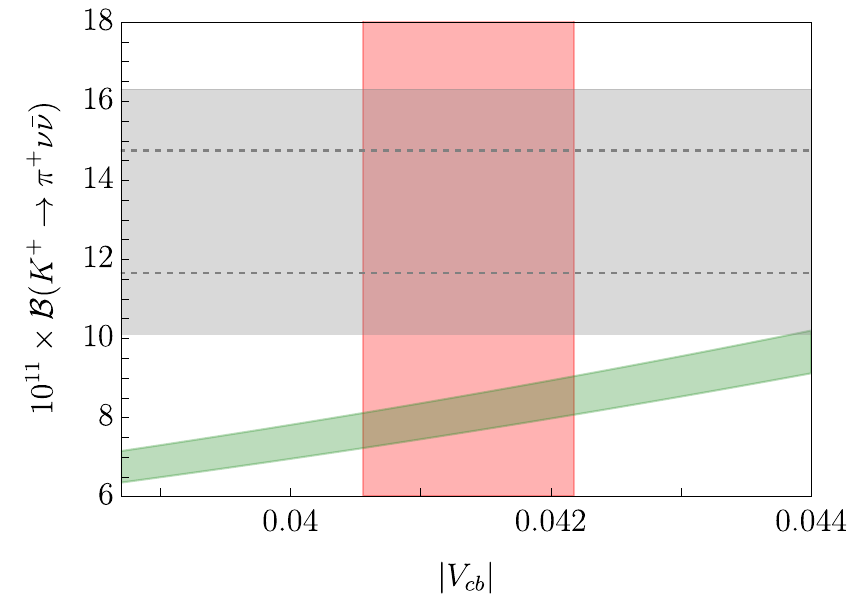}
\caption{Parametric dependence of $\cB(K^+\to\pi^+\nu\bar\nu)$ on $|V_{cb}|$, within the SM (green band).
The red band denotes the current $|V_{cb}|$ value in Eq.~(\ref{eq:Vcb_combination}).
The gray band indicate the $1\sigma$ interval of the current experimental measurement of 
$\cB(K^+\to\pi^+\nu\bar\nu)$~\cite{NA62talk}, 
while the dotted lines denote the projected NA62 uncertainty at the end of Run 3.}
\label{Fig:KtopivsVcb}
\end{figure}

 Despite the many efforts in the recent years, the exclusive and inclusive determinations of $|V_{cb}|$ are still in tension. In particular, a lower value of $|V_{cb}|$, as the exclusive determinations suggest, would push the predictions for these two decay modes towards lower values. We opt for the following strategy: we combine the global fit for the inclusive determination \cite{Finauri:2023kte} together with the recent exclusive one in \cite{Bordone:2024weh}\footnote{The exclusive determination in \cite{Bordone:2024weh} agrees with previous analyses in \cite{Martinelli:2021onb,Martinelli:2021myh,Martinelli:2022xir,Martinelli:2023fwm} and with the global fits in \cite{UTfit:2022hsi,UTfit:ICHEP,FlavourLatticeAveragingGroupFLAG:2024oxs}.}. This leads to
\begin{equation}
    |V_{cb}|_\mathrm{incl+excl} =\, (41.37 \pm 0.81)\times 10^{-3}\,,
    \label{eq:Vcb_combination}
\end{equation}
where the uncertainty has been inflated to account for the tension between the two determinations. We can see the dependence of the $K^+\to \pi^+\nu\bar\nu$ branching fraction on $|V_{cb}|$ in Fig.~\ref{Fig:KtopivsVcb}. Here the green band is the parametric dependence of the theoretical prediction from $|V_{cb}|$, the gray band is the up-to-date experimental measurement and in red we represent the value for $|V_{cb}|$ in Eq.~(\ref{eq:Vcb_combination}). From the plot it is clear how a lower $|V_{cb}|$ value would increase the tension with the SM prediction with respect to our choice in Eq.~(\ref{eq:Vcb_combination}). This problem will become even more evident in the next few years, in view of 
projected NA62 uncertainty on $\cB(K^+\to\pi^+\nu\bar\nu)$
of $\sim 15\%$ at the end of Run 3.

As pointed out in Ref.~\cite{Buras:2021nns,Buras:2022qip},
one can eliminate the $|V_{cb}|$ dependence of $\cB(K^+\to\pi^+\nu\bar\nu)$ constructing appropriate ratios with $\Delta F=2$ observables (in particular $\epsilon_K$). However, given the latter are affected by NP of different dynamical origin with respect to the one relevant to $K\to\pi\nu\bar\nu$, this complicates the subsequent NP analysis. 
Our scope is to obtain a SM prediction of $\cB(K\to\pi\nu\bar\nu)$ which is both conservative and based on SM inputs which are not affected by NP (at least within our framework).

To this purpose, we determine the CKM parameters in the following way. We fix $\bar\rho$ and $\bar{\eta}$ using the UTfit global fit to observables sensitive to the angles only \cite{UTfit:2022hsi,UTfit:ICHEP}. We then extract $\lambda$ from super-allowed $\beta$ decays, and finally use the $|V_{cb}|$ value in 
(\ref{eq:Vcb_combination}) to 
determine $A\doteq |V_{cb}|/\lambda^2$. 
Proceeding this way we obtain 
\begin{equation}
    \begin{aligned}
        \lambda =&\,0.2251\pm 0.0008\,,  &  \bar\rho =&\, 0.144 \pm 0.016\,,  \\
        A =&\, 0.816 \pm 0.017\,,  &  \bar\eta =&\,  0.343 \pm 0.012\,.
    \end{aligned}
\end{equation}
With these inputs, we predict
\begin{align}
    \mathcal{B}(K^+\to \pi^+\nu\bar\nu)^{\rm SM}=&\, (8.09 \pm 0.63)\times 10^{-11}\,,\label{K+p+}\\
    \mathcal{B}(K_L\to \pi^0\nu\bar\nu)^{\rm SM}=&\,(2.58 \pm 0.30)\times 10^{-11}\,,
\end{align}
that we employ as the reference SM values in the rest of this work.

\section{EFT framework}
\label{sect:EFT}
As anticipated, we work within an EFT framework assuming heavy NP predominantly coupled to third-generation fermions. More precisely, we assume that the only dynamical fields are the SM ones, and we neglect $U(2)^5$--invariant operators involving light fermions. On the other hand, since we are interested in 
 describing flavor mixing in the quark sector, we consider operators built in terms of the leading $U(2)_q$--breaking spurion $\tilde{V}$~\cite{Faroughy:2020ina} 
which is responsible for the heavy$\to$light mixing in the quark 
Yukawa couplings. The spurion, which transforms as a doublet under  $U(2)_q$,
is parameterized as
\begin{equation}
  \tilde{V}= -\varepsilon V_{ts}  \begin{pmatrix}
     \kappa V_{td}/V_{ts} \\[1.5mm]
     1
    \end{pmatrix}\,.
    \label{eq:Vform}
\end{equation}
The parameters $\varepsilon$ and $\kappa$ are assumed to be real and $O(1)$:
$\varepsilon$ control the overall size of the spurion  (the normalization is chosen such that $3\to 2$ mixing is positive for
${\varepsilon>0}$ in the standard CKM convention),
while $\kappa$ quantifies possible  deviation from a minimal $U(2)_q$--breaking
structure. The minimal framework 
corresponds to the  limit $\kappa=1$~\cite{Barbieri:2011ci,Faroughy:2020ina}.

In this setup there is an intrinsic ambiguity on what we denote as third generation 
in the left-handed quark sector, or better 
which are the $U(2)_q$ singlet fields. 
For definiteness, we choose a down-aligned basis, where the quark doublets are written as $q_L^i = (V_{ji} u_L^j, d_L^i)^T$, with $u_L^i$ and $d_L^i$ denoting the quark mass eigenstates, such that 
\begin{equation}
q_L^3= \begin{pmatrix}
    V_{ub} u_L +  V_{cb} c_L +  V_{tb} t_L \\
    b_L 
    \end{pmatrix}
    \label{eq:q3}
\end{equation}
is a $U(2)_q$ singlet.
On the lepton side, we focus only on amplitudes sensitive to third-generation
leptons, hence the only relevant lepton fields to consider are $\ell_L^3 = (\nu_{\tau}, \tau_L)^T$ and $\tau_R$. With these assumptions, the leading semileptonic operators involving only 
$U(2)^5$--singlet  fields are 
\begin{align}
Q_{\ell q}^{\pm}&=(\bar{q}_L^3 \gamma^{\mu} q_L^3)(\bar{\ell}_L^3\gamma_{\mu}\ell_L^3)  \pm ( \bar{q}_L^3 \gamma^{\mu}\sigma^a q_L^3)(\bar{\ell}_L^3\gamma_{\mu}\sigma^a\ell_L^3)\,,
\nonumber \\[1.3mm]
Q_{S}&=(\bar{\ell}_L^{3}\tau_R)(\bar{b}_R q_L^3)\,.
\label{eq:ops}
 \end{align}
 The Wilson Coefficients associated to these three operators are dimensionful parameters that we express in units of $\mathrm{TeV}^{-2}$:
\begin{equation}\label{eq:lagrangian}
\mathcal{L}^{\rm NP}_{\mathrm{eff}}\supset \sum_k C_k Q_k+{\rm h.c.}\,.
\end{equation}

In principle, in addition to the three operators 
in Eq.~(\ref{eq:ops}) we should consider all terms generated by the insertion of one or two spurions in each of them, separately, via the replacement $q_L^3 \to \tilde{V}_i q_L^i$. 
In practice, to avoid the proliferation of free parameters, we assume that the underlying NP leads to a rank-one structure in quark flavor space~\cite{Marzocca:2024hua}.
In other words, we assume that NP is aligned to a specific direction in flavor space, and the insertion of spurions describes the misalignment of this direction relative to that of the $q_L^3$ field
in (\ref{eq:q3}). In practice, this condition is achieved via the replacement 
\begin{equation}
    q_L^3 \to q_L^3 + \tilde{V}_i q_L^i
\end{equation}
in the three operators in Eq.~(\ref{eq:ops}).
In Sect.~\ref{sect:numerics} we will discuss what are the implications of the rank-one hypothesis, relative to the more general case, and we will provide explicit examples of  ultraviolet  (UV) completions where this condition is fulfilled.

The EFT framework we are considering is then
described by five independent parameters: $C_{\ell q}^{+}$, $C_{\ell  q}^{-}$, $C_S$, $\varepsilon$, and $\kappa$. 
The list of the observables included in the analysis and the way they are affected by these parameters is summarized in Table \ref{tab:obs}.

\begin{table}[t]
\centering
\renewcommand{\arraystretch}{1.5} 
\begin{tabular}{|c|c c c c c || c |}
\cline{2-7}
\multicolumn{1}{c|}{} & $C_S$ & $C_{\ell q}^{+}$ & $C_{\ell q}^{-}$ & $\varepsilon$ & $\kappa$ & 
{\rm Exp.~indication} \\
\hhline{-======}
$\sigma(pp \to \ell\ell)$  & $\checkmark$ & $\checkmark$ & $\checkmark$ &   & & 
bounds~on~$\cA_{\rm NP}$ \\
\hline
EWPO  &  & $\checkmark$ & $\checkmark$ &  & & 
bounds~on~$\cA_{\rm NP}$ \\
\hline\hline
$R_D$, $R_{D^*}$ & $\checkmark$ & $\checkmark$ & $\checkmark$ & $\checkmark$ &  & 
$\cA_{\rm NP}/\cA_{\rm SM} > 0 $  \\
\hline
$ \cB(B \to K^{(*)} \mu\bar\mu)$  &  & $\checkmark$ &  & $\checkmark$ & &  
$\cA_{\rm NP}/\cA_{\rm SM} < 0 $  \\
\hline\hline
$ \cB(B \to K \nu\bar\nu)$ &  &  & $\checkmark$ & $\checkmark$ &  &
~$|\cA_{\rm SM}+\cA_{\rm NP}|^2 > |\cA_{\rm SM} |^2$  \\
\hline
$ \cB(K \to \pi \nu\bar\nu)$  &  &  & $\checkmark$ &  & $\checkmark$ &  ~$|\cA_{\rm SM}+\cA_{\rm NP}|^2 > |\cA_{\rm SM} |^2$  \\
\hline
\end{tabular}
\caption{List of observables considered in the analysis, and their sensitivity to the EFT parameters. In the last column we highlight the present hints of deviations from the SM, as emerging from data.}
\label{tab:obs}
\end{table}

\subsection{Flavor-changing amplitudes}
Before presenting the numerical results, it is useful to discuss the implications of the $U(2)_q$ breaking assumptions on the different 
flavor-changing
amplitudes. While the complete EFT predictions are illustrated in detail in the appendix, here we provide some simplified formulae which illustrate the main effects.  

Let's start from the contributions to $R_{D}$ and $R_{D^*}$. Here NP interfere with a tree-level SM amplitude, hence can be treated as a small correction. Considering only the vector operators in (\ref{eq:ops})
and expanding to first order in their coefficients leads to
\begin{align}
    \frac{R_{D^{(*)}}}{R_{D^{(*)}}^{\textrm{SM}}} &\approx 1 + 2 {\rm Re}\left( C_{V_L} \right) 
    \nonumber \\
    &\approx 1  - v^2 \left(1+\varepsilon\right) \left(C_{\ell q}^{+}-C_{\ell q}^{-}\right)\,,
    \label{eq:RDsimp}
\end{align}
where $v= (\sqrt{2} G_F)^{-1/2} \approx 246$~GeV, the explicit expression of $C_{V_L}$
can be found in (\ref{eq:CVL})
and we have neglected sub-leading terms of $O(\lambda^2)$.
As can be noted, the observable receives a non-vanishing correction also in absence of spurion contributions ($\varepsilon\to 0$ limit), while the effect of the spurion is constructive for $\varepsilon >0$.
As is well known,  the vector operators
$Q_{\ell q}^{\pm}$ lead to a universal shift in $R_{D}$ and $R_{D^*}$, only the scalar operator differentiate among the two observables (see appendix).

As far as FCNC processes are concerned, it is easy to realize that  $b\to s$ transitions are sensitive to the insertion of a single spurion (NP amplitude 
proportional to $\varepsilon$), while $s\to d$ modes are affected by the insertion of two spurions,  with an amplitude proportional to $\varepsilon^2\kappa$. In the neutrino modes,  NP modifies (at the tree level) the coefficients of the effective Lagrangian in Eq.~(\ref{eq:Leff})
involving tau neutrinos:
\begin{equation}
\label{eq:nnmodes}
    \begin{split}
    |C_{\tau,bs}^{\textrm{SM}} |
      &\to \left| C_{\tau,bs}^{\textrm{SM}}
      -\varepsilon \frac{\pi v^2}{\alpha} 
      C_{\ell q}^{-}\right|\,,\\
     |C_{\tau,sd}^{\textrm{SM}} |
      &\to \left| C_{\tau,sd}^{\textrm{SM}}
      + \kappa\varepsilon^2 \frac{\pi v^2}{\alpha} 
      C_{\ell q}^{-}\right|\,.
    \end{split}
\end{equation}
The effect is potentially sizable, given the SM amplitude is loop suppressed. Note that, since 
$C_{\tau,bs}^{\textrm{SM}} \approx 
C_{\tau,sd}^{\textrm{SM}}$, the
interference between SM and NP amplitude is opposite in the 
two modes in the limit $\kappa=1$ and $\varepsilon>0$. 
Since in our framework there are no scalar nor right-handed current
operators affecting the di-neutrino modes, a firm prediction of this setup
is a universal  modification of
$\cB(B^+ \to K^+\nu\bar\nu)$ and $\cB(B\to K^*\nu\bar\nu)$
relative to their SM values.
This differs from what happens 
in  more general beyond-SM  frameworks~\cite{Bause:2023mfe,Allwicher:2023xba,Buras:2024ewl}.

The last flavor-violating effect we consider is the modification of the $b\to s \ell\bar\ell$ amplitude
($\ell=e,\mu$) which occurs from 
the QED running of the $(\bar b_L\gamma^\mu s_L) (\bar \tau_L \gamma_\mu \tau_L)$ 
operator into $Q_9 \propto  (\bar b_L\gamma^\mu s_L) (\bar \ell \gamma_\mu \ell)$~\cite{Aebischer:2017gaw,Crivellin:2018yvo}. This results into a lepton-universal shift of the Wilson coefficient~$C_9$, defined as in~\cite{Aebischer:2017gaw}, which can be written as~\cite{Aebischer:2022oqe} 
\begin{equation}
\frac{ C^{\rm NP}_9 }{C_9^{\rm SM}} 
\approx  \frac {\varepsilon  v^2 }{3}  \frac{ C_{\ell q}^{+}}{C_9^{\rm SM} } 
\log{\left(
    \frac{\Lambda^2}{m^2_\tau}\right)}
    \approx 0.23 \times \varepsilon C_{\ell q}^{+}
    \,.
\end{equation}
The numerical expression has been obtained setting $\Lambda =1$~TeV
(see the appendix for more details). As for $R_{D^{(*)}}$, also in this case the NP effect can at most be a small correction with the respect to the SM, given both NP and SM amplitudes are loop suppressed.

\begin{figure}[t]
\includegraphics[width =8.0cm]{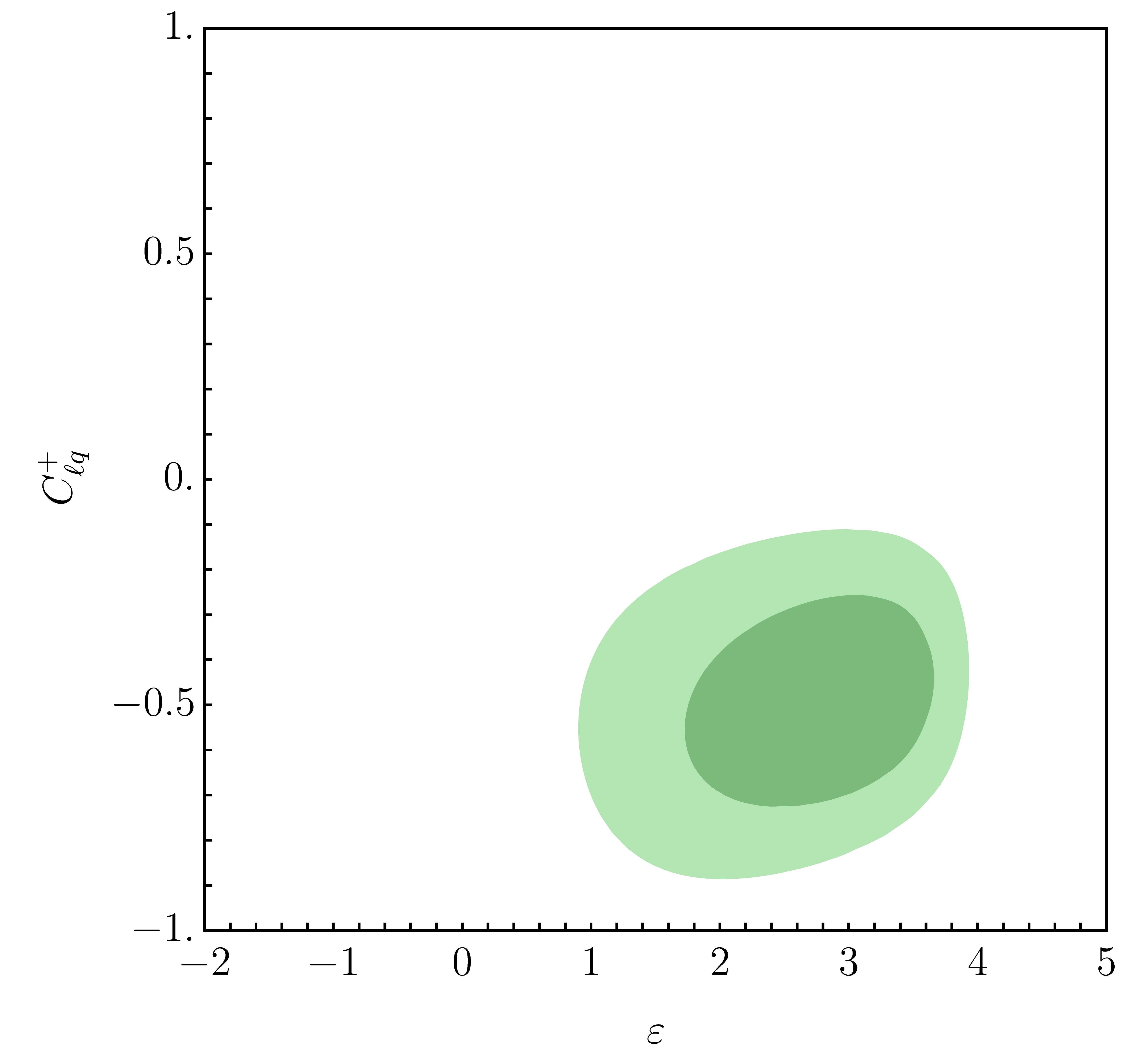}
    \caption{\label{Fig:Cp} Constraints in the $C_{\ell q}^{+}$--$\varepsilon$ plane.
    The green areas denote the parameter regions favored at 1$\sigma$ and 2$\sigma$ from a fit to all the observables but the two di-neutrino rare modes.}
\end{figure}

\begin{figure}[t]
 \includegraphics[width = 8.0cm]{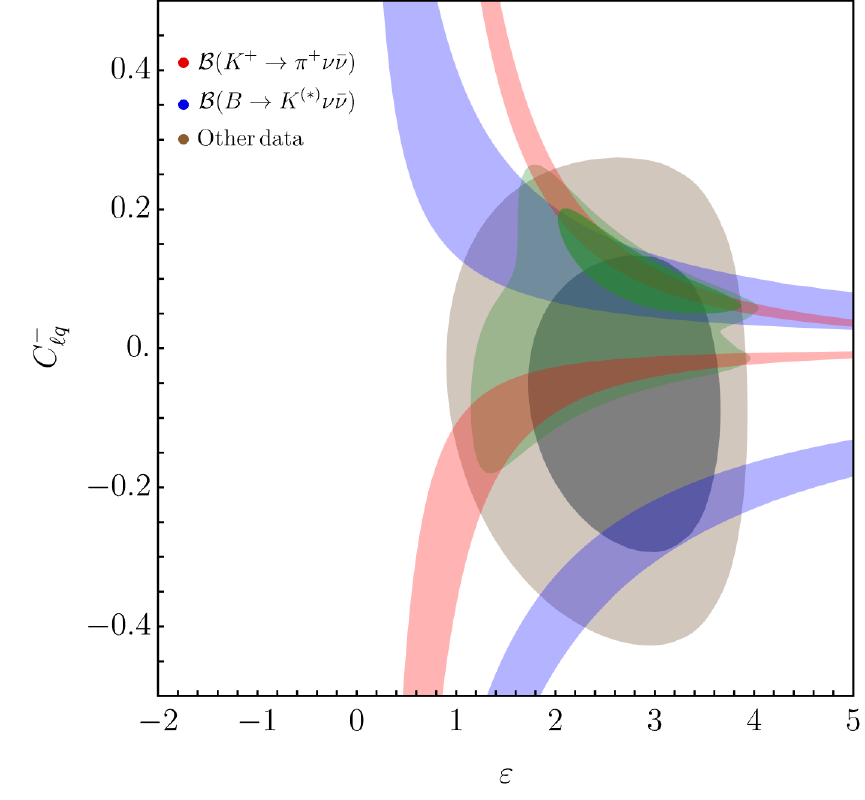} 
    \caption{\label{Fig:Cm} Constraints in the $C_{\ell q}^{-}$--$\varepsilon$ plane.
    The brown areas denote the parameter regions favored at $1\sigma$ and $2\sigma$ from a fit to all the observables but di-neutrino modes. The regions favored by  
    $\cB(B \to K^{(*)} \nu\bar \nu)$ and $\cB(K^+ \to \pi^+ \nu\bar \nu)$, separately, in the limit $\kappa=1$, are indicated in blue and red, respectively ($1\sigma$ bands). The green areas are the regions favored by all data at $1\sigma$
    and $2\sigma$
    ($1\sigma$ only around the best fit point).}
\end{figure}

\section{Numerical Analysis}
\label{sect:numerics}

We now proceed with a combined numerical analysis of the different observables listed in Table \ref{tab:obs}. We can distinguish three classes of observables: i)~EWPO and direct searches, which provide stringent bounds on $C_{\ell q}^\pm$
and $C_S$; ii)~$R_{D}$, $R_{D^*}$ and $ \cB(B \to K^{(*)} \mu\bar\mu)$, where NP is at most a small correction with respect to the SM; iii) the two neutrino modes, where $O(1)$ modifications of the SM amplitude are possible. 

There is a partial ``factorization" of the different constraints: observables i) and ii) provide a stringent determination of $C_{\ell q}^{+}$ and $\varepsilon$, as illustrated in Fig.~\ref{Fig:Cp}. Obviously, the preference for non-vanishing values of  $C_{\ell q}^{+}$ and $\varepsilon$ is driven by the set ii).
In particular, a positive value of $\varepsilon$ is favored by Eq.~(\ref{eq:RDsimp}), since it allows a constructive interference between spurion and non-spurion terms in $R_{D}$ and $R_{D^*}$, hence maximal contribution at fixed $C_{\ell q}^{\pm}$, whose absolute size is constrained by set i). The maximal value of $\varepsilon$ is not determined by data but by model-dependent considerations: since we assume 
$\varepsilon=O(1)$, we have added a theoretical likelihood to suppresses values of $|\varepsilon|$ above $3$.
Scalar contribution do not alter this picture, being strongly constrained by direct searches (the best fit value of $C_S$ is compatible with zero). 

The two di-neutrino modes have a marginal role in constraining $C_{\ell q}^{+}$ and $\varepsilon$. On the other hand, despite their sizable experimental uncertainty, they already provide a significant constraint on the value of $C_{\ell q}^{-}$.
As shown in Fig.~\ref{Fig:Cm}, the parameter $C_{\ell q}^{-}$ is largely unconstrained by the observables in sets i) and ii), beside its maximal allowed size, while the information from the two di-neutrino modes provides a relevant constraint. 
Note that in 
Fig.~\ref{Fig:Cm} the band denoted $\cB(B\to K^{(*)} \nu\bar \nu)$
corresponds to the combined constraint from  $\cB(B^+\to K^+ \nu\bar \nu)$ and $\cB(B\to K^* \nu\bar \nu)$, which in our framework have the same functional dependence.

\begin{figure}[t]
\centering
\includegraphics[width = 7.0cm]{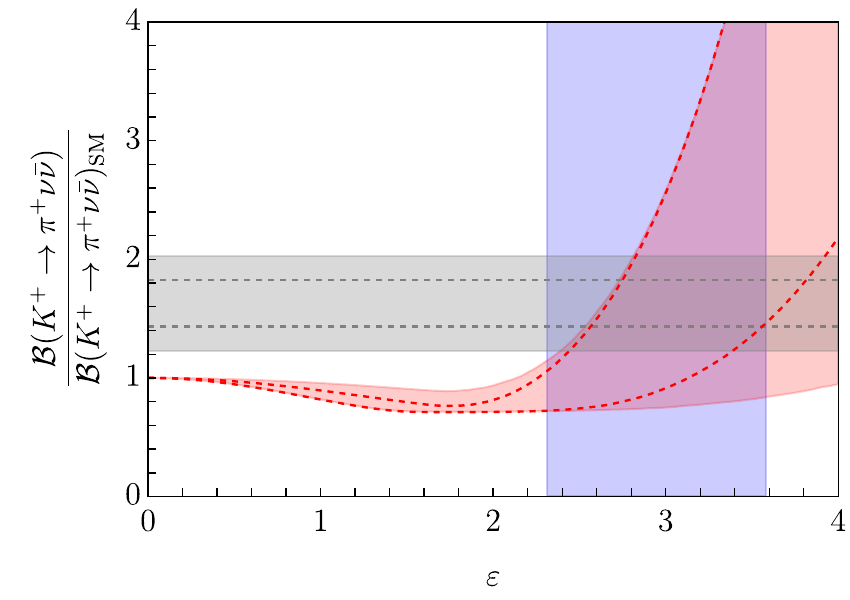} 
    \caption{\label{Fig:Kpnnvse}
    Prediction of $\cB(K^+\to \pi^+\nu\bar\nu)$ as function of $\varepsilon$, for $\kappa=1$, using the value of $C_{\ell q}^-$ determined from all the other observables at $1\sigma$ (red band).
     The blue band indicates the preferred $\varepsilon$ values ($1\sigma$) from the global fit without $\cB(K^+\to \pi^+\nu\bar\nu)$.
     The gray band indicates the experimental determination of $\cB(K^+\to \pi^+\nu\bar\nu)$ at $1\sigma$.
     The red and gray dashed lines illustrate the change of the respective regions assuming near-future experimental projections for  $\cB(K^+\to \pi^+\nu\bar\nu)$ and $\cB(B\to K^{(*)}\nu\bar\nu)$
     (see text). The red dashed lines are not centered with respect to the red band given the central value of $C_{\ell q}^-$ changes with the projected branching fractions. We do not include a future projection on the $\varepsilon$ (blue) band since this is not controlled only by the di-neutrino modes.} 
\end{figure}

In the limit of minimal $U(2)_q$ breaking, i.e.~setting ${\kappa=1}$, there is good compatibility of the constraints from the two rare modes already at $1\sigma$, in the parameter region for $\varepsilon$ favored by the other data. This is non trivial given the decrease of $\cB(K^+\to \pi^+\nu\bar\nu)$, compared to its SM value, for 
$0 < \varepsilon < 2$,  as indicated in Fig.~\ref{Fig:Kpnnvse}.   When all data are combined, a positive value of $C_{\ell q}^{-}$ is preferred. 

\begin{figure}[t]
\includegraphics[width = 7.3cm]{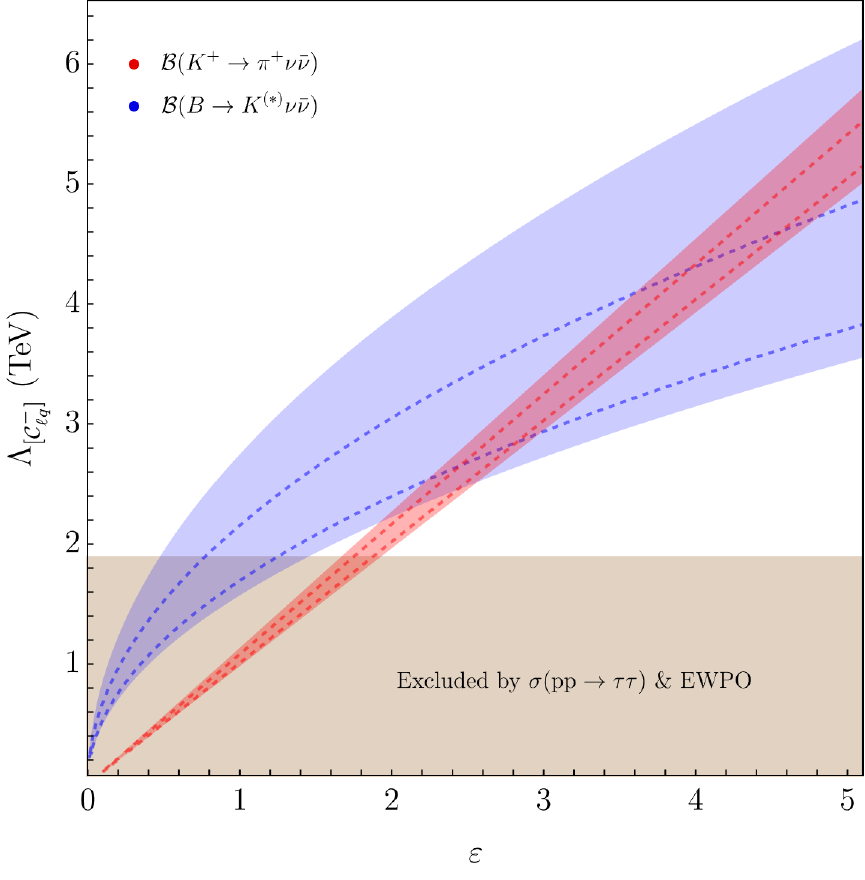} 
    \caption{\label{Fig:Lambdafig}
    $\Lambda$ vs.~$\varepsilon$ plane, where $\Lambda$  is defined by $C_{\ell q}^- = 1/\Lambda^2$. The blue and red areas indicate  present $1\sigma$ constraints from  $\cB(K^+\to \pi^+\nu\bar\nu)$ and $\cB(B\to K^{(*)}\nu\bar\nu)$, setting $\kappa=1$. The 95\%~CL exclusion limit from direct searches and EWPO are also indicated. The  dashed lines illustrate the change of the respective regions assuming near-future experimental projections for  $\cB(K^+\to \pi^+\nu\bar\nu)$ and $\cB(B\to K^{(*)}\nu\bar\nu)$. }
\end{figure}

\subsection{Discussion}

From the numerical results presented above we can derive the following conclusions.

\begin{itemize}
\item 
The factorization of the constraints implies that the rank-one hypothesis we have implemented is not really tested by present data. As far as the observables i) and ii) are concerned, the only relevant spurion term is $Q_{\ell q}^+$ (with one spurion):  a generic coupling for this operator can thus be reabsorbed in the value of $\varepsilon$. As far as the two di-neutrino modes are concerned, possible $O(1)$ couplings in the relevant operators can be reabsorbed in the values of $C_{\ell q}^-$ and $\kappa$. In other words, given current precision, five effective parameters are necessary to describe the system in full generality given the main dynamical assumptions of a mildly broken $U(2)^5$ flavor symmetry. 
\item
A non trivial test we can perform with present data is the viability of the hypothesis of a minimal breaking of $U(2)_q$, combined with the rank-one relation between $\cB(B\to K^{(*)}\nu\bar\nu)$ and $\cB(K^+\to \pi^+\nu\bar\nu)$. Figs.~\ref{Fig:Cm}--\ref{Fig:Kpnnvse} show that this hypothesis, which corresponds to fix $\kappa=1$ and reduce the number of free parameters from 5 to 4, is well supported by present data. 
A further reduction from 4 to 3 parameters, obtained setting $C_S=0$, is also well supported by data. 
\item 
The two di-neutrino modes test scales well above those directly probed at colliders. A clear illustration of this fact is shown in Fig.~\ref{Fig:Lambdafig},
where the value of $C_{\ell q}^-$ is expressed in term of the corresponding  effective scale $\Lambda$. 
In principle, combining two modes could lead to a determination of $\varepsilon$ completely independent from that obtained from the other flavor-changing observables. Right now this is not very significant given the large experimental uncertainties. Note that the study of normalised kinematic distributions for the di-neutrino modes would not add additional information since in our setup we generate only SM-like effective operators below the electroweak scale (see \cite{Gorbahn:2023juq,Buras:2024ewl} for non-SM like examples).

\begin{figure}[t]
\includegraphics[width = 8cm]{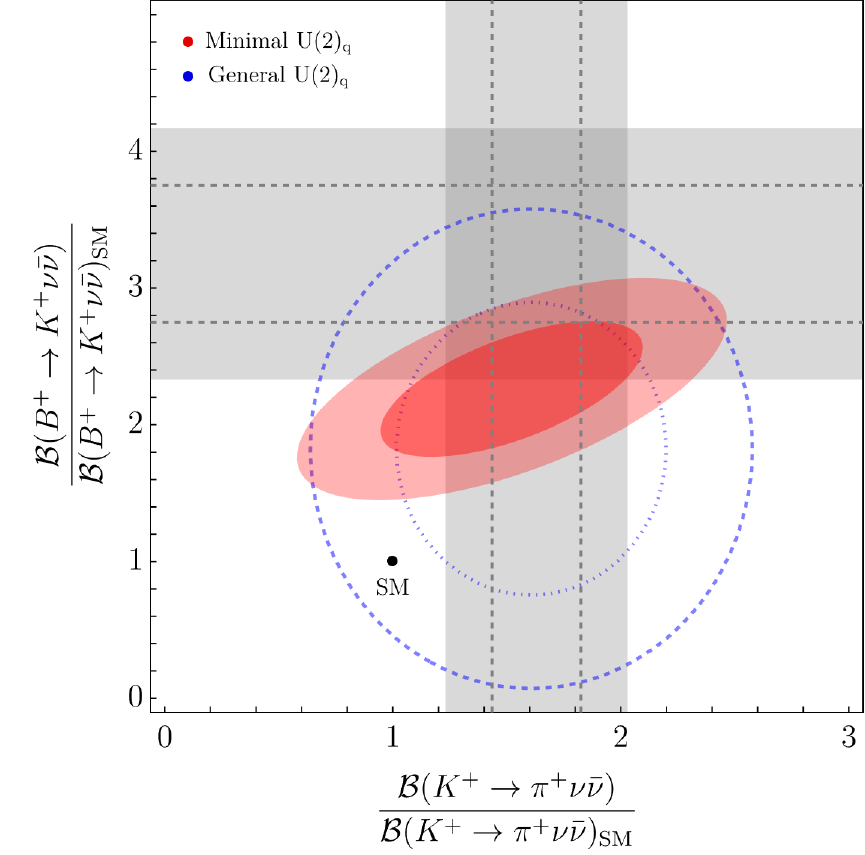} 
    \caption{\label{Fig:BvsK}
    Correlation between $\cB(B^+\to K^+\nu\bar\nu)$ and $\cB(K^+\to \pi^+\nu\bar\nu)$, normalized to their SM predictions.
    The red areas denote the parameter regions favored at $1\sigma$ and $2\sigma$ from a global fit in the limit of minimal $U(2)_q$ breaking (${\kappa=1}$).
    The dashed and dotted blue curves are $1\sigma$ and $2\sigma$ regions from a global fit where $\kappa$ is a free parameter. The gray bands indicate the current experimental constraints, while the dashed gray lines highlight near-future projections assuming halved
    experimental uncertainties. }
\end{figure}

\item
Assuming minimal $U(2)_q$ breaking, the combination of the two di-neutrino modes leads to a $2\sigma$
hint of NP. This is evident in Fig.~\ref{Fig:BvsK}, where we show the favored region in the 
$\cB(B^+\to K^+\nu\bar\nu)$--$\cB(K^+\to \pi^+\nu\bar\nu)$ plane from the global fit.
As also shown in Fig.~\ref{Fig:BvsK}, relaxing the assumption of minimal $U(2)_q$ breaking the two modes become uncorrelated and the combined significance for a deviation from the SM drops below 95\% CL.\footnote{The small tension of the results of the global fits with the experimental determination of 
$\cB(B^+\to K^+\nu\bar\nu)$ is caused by the inclusion of 
$\cB(B\to K^*\nu\bar\nu)$ data in the global fit.
}
\item
In Figs.~\ref{Fig:Kpnnvse}--\ref{Fig:BvsK} we illustrate how the individual constraints on the two modes, or the respective measurements, could change in the near future, assuming same central values but halved experimental errors. As can be seen in Fig.~\ref{Fig:Lambdafig}, this would allow us to derive a stringent range for $\varepsilon$, testing the overall consistency of this framework and, in particular, the validity of the assumption $\varepsilon=O(1)$. 
As already stated, a firm prediction of this framework is a relative deviation from the SM identical in 
$\cB(B^+\to K^+\nu\bar\nu)$ and $\cB(B\to K^*\nu\bar\nu)$, we thus expect the central value of  $\cB(B^+\to K^+\nu\bar\nu)$ to decrease, as indicated in Fig.~\ref{Fig:BvsK}.
If the central value of $\cB(K^+\to \pi^+\nu\bar\nu)$ would remain unchanged, the combination of the two modes 
under the hypothesis of  minimal $U(2)_q$ breaking, as in Fig.~\ref{Fig:BvsK}, would point toward a deviation from the SM well above the $3\sigma$ level.
\end{itemize}

\subsection{Comparison with explicit models}

\paragraph{Vector leptoquark.}
The non-vanishing values of $C_{\ell q}^\pm$ and 
$\varepsilon$  are qualitatively in good agreement with the effects associated to 
a TeV-scale vector leptoquark coupled mainly to the third-generation
(see e.g.~\cite{Barbieri:2015yvd, DiLuzio:2017vat, Fuentes-Martin:2020hvc}). While the values of $C_{\ell q}^+$ and $\varepsilon$ confirm previous findings along this direction~\cite{Aebischer:2022oqe},  the di-neutrino modes provide an additional support to this picture indicating $|C_{\ell q}^+| \gg |C_{\ell q}^-|$. This is expected given 
$|C_{\ell q}^+|$ arises by the tree-level exchange of the  leptoquark, while
$|C_{\ell q}^-|$ is generated at the loop level~\cite{Fuentes-Martin:2020hvc}. 

It is worth stressing that 
both magnitudes and signs of the NP amplitudes are consistent with predictions made, before the observations of the two decay modes,
in complete models where the leptoquark is 
a massive gauge boson arising from 
$SU(4)^{[3]}\times SU(3)^{[12]} \to SU(3)_{c}$~\cite{Fuentes-Martin:2019ign,Fuentes-Martin:2020luw,
Fuentes-Martin:2020hvc, Crosas:2022quq}.   In this case, 
the constructive interference between SM and NP amplitudes in $B\to K^{(*)} \nu\bar\nu$ is unambiguously related to the enhancement of $R_{D^{(*)}}$, as shown 
first in~\cite{Fuentes-Martin:2020hvc} (based on the results in~\cite{Fuentes-Martin:2019ign,Fuentes-Martin:2020luw}).

The $K\to \pi\nu\bar\nu$ amplitude is more complicated, since non-minimal $U(2)_q$ terms are naturally present; however, as shown in \cite{Crosas:2022quq}, constraints from $\Delta S=2$ amplitudes point to a scenario that, once expressed in our notation, effectively corresponds to $1.0\lesssim {\rm Re}(\kappa) \lesssim{1.5} $.

A discussion of the correlations between $K$ and $B$ decays under the $U(2)$ hypothesis in the scalar leptoquark scenario can be found in \cite{Marzocca:2021miv}.

\paragraph{$Z^\prime$ boson.}
A generic $Z^\prime$ boson coupled to left-handed  quarks and leptons provides a useful example to illustrate some of the challenges in generating large corrections to the di-neutrino widths in explicit models.  Consider a massive $Z^\prime$ coupled to the following current 
\begin{equation}
J^\mu_{Z^\prime} = Q_q (\bar{q}_L^3 + \tilde{V}^*_i \bar q_L^i)
\gamma^\mu (q_L^3 + \tilde{V}_i q_L^i)
+ Q_\tau \bar{\ell}_L^3\gamma^\mu \ell_L^3\,.
\end{equation}
Integrating out the $Z^\prime$ leads to 
\begin{align}
C_{\ell q}^{-} =
C_{\ell q}^{+} &= - \frac{g^2}{M_{Z^\prime}^2} Q_q Q_\tau\,, 
\label{eq:CmZ}\\
C_{qq}^{(1)[3333]} &= -
\frac{g^2}{2 M_{Z^\prime} ^2} Q_q^2 \,.
\label{eq:DF2}
\end{align}
The term in~(\ref{eq:DF2}) is the coefficient of the four-quark operator, in the notation of Ref.~\cite{Allwicher:2023shc}. This is severely bounded by $\Delta F=2$ amplitudes: generalising the up-aligned result of~\cite{Allwicher:2023shc} (which corresponds to setting $|\varepsilon|=1$ in our case), leads to 
\begin{equation}
\big|C_{qq}^{(1)[3333]} \big| < (\varepsilon \Lambda_{Bs})^{-2}\,,
\end{equation}
with $\Lambda_{Bs} = 7.6$~TeV. On the other hand, from Fig.~\ref{Fig:Lambdafig} we deduce that fitting the central values of the two di-neutrino modes requires 
\begin{equation}
   \big|C_{\ell q}^{-} \big| \approx \left[\varepsilon \times (1~{\rm TeV})\right]^{-2}\,.
   \label{eq:CmZexp}
\end{equation}
Satisfying Eqs.~(\ref{eq:CmZ})--(\ref{eq:CmZexp}) is possible, but only if $|Q_\tau/Q_q|\gtrsim 30$, which appears a rather tuned choice. 
This excludes, for instance, the $Z^\prime$ bosons arising from flavor deconstruction of $U(1)$ gauge groups~\cite{
FernandezNavarro:2023rhv,Davighi:2023evx,Barbieri:2023qpf}. 
The difference with respect to the leptoquark case discussed above, also arising from flavor deconstruction, is that deconstructing the $SU(4)$
group one can conceive a flavor misalignment of charged currents (leptoquark-mediated) and neutral currents ($Z^\prime$-mediated), avoiding the strong tree-level bounds from $\Delta F=2$ transitions~\cite{DiLuzio:2018zxy}.

\section{Conclusions}\label{sect:conclusions}

The recent observation of the $K^+\to \pi^+ \nu\bar\nu$
decay by NA62~\cite{NA62talk}, and the evidence of $B^+\to K^+ \nu\bar\nu$ reported by Belle-II~\cite{Belle-II:2023esi}, 
represent a major step forward in flavor physics. 
These long-sought rare decays are unique probes of short-range dynamics and,  as pointed out in a number of recent works~\cite{Marzocca:2024hua,Buras:2024ewl,Bause:2023mfe,Allwicher:2023xba,Anzivino:2023bhp,Bause:2021cna}, they are sensitive to a variety of SM extensions. 
Given the two neutrinos cannot be detected, also extensions of the SM  with exotic invisible light states may affect these decay modes (see e.g.~\cite{Altmannshofer:2023hkn,McKeen:2023uzo,
Fridell:2023ssf,Gabrielli:2024wys,He:2024iju,Bolton:2024egx,Kim:2024tsm,Hou:2024vyw,He:2024iju,Hou:2024vyw}).

In this paper, we have analyzed these processes in a specific class of SM extensions with heavy new particles.
More precisely, we have considered an EFT based on the hypothesis of TeV-scale NP coupled mainly to quarks and leptons of the third generation. 
On general grounds, this hypothesis is particularly interesting for addressing both the hierarchy problem and the flavor hierarchies~\cite{Allwicher:2023shc,Davighi:2023iks} . Moreover, it is a setup where sizable deviations from the SM can consistently occur in the rare di-neutrino modes, despite the tight NP bounds derived from $B_s\to \mu^+\mu^-$ and other flavor-changing processes.
Indeed, as already stated in the introduction, $K\to \pi \nu\bar\nu$ and $B\to K^{(*)} \nu\bar\nu$ are the 
only FCNC transitions involving third-generation leptons for which current measurements have reached the SM level.

In the motivated NP framework we have considered, $K\to \pi \nu\bar\nu$ and $B\to K^{(*)} \nu\bar\nu$ amplitudes are naturally linked. The link is non trivial and provides a key information to determine the flavor structure of the underlying dynamics. 
Despite affected by a significant uncertainty, present data hint to an enhancement of both decay rates compared to the corresponding SM expectations. As we have shown, 
this hint is  compatible with being generated by a unique effective operator, with an alignment in flavor space following from the hypothesis of a minimally broken $U(2)_q$ flavor symmetry. This flavor structure is also fully compatible, and naturally linked, with other hints of deviations from the SM in $B\to D^{(*)}\tau\nu$ and $B \to K^{(*)} \mu\bar \mu$ decays and, at the same time, is compatible with bounds from electroweak physics and direct searches.
Enhanced rates for the two di-neutrino modes were indeed predicted~\cite{Fuentes-Martin:2020hvc,Crosas:2022quq}, before the recent measurements~\cite{NA62talk,Belle-II:2023esi}, in complete UV models addressing these other observations.
Needless to stress that future data on all these modes are needed to understand how solid this picture is. In the case of $B\to D^{(*)}\tau\nu$ and $B \to K^{(*)} \mu\bar \mu$,  further theoretical scrutiny of the SM uncertainties is also needed. 

Given their theoretical cleanliness, the di-neutrino modes plays a special role in shedding more light on this framework. As we have shown, already a reduction of a factor of two of present uncertainties could bring the evidence of NP, from the di-neutrino modes only,
above the $3\sigma$ level (if the overall picture would not change significantly). More generally, the analysis we have presented provides a clear illustration of the high discovery potential  and the unique discriminating power of these rare modes, especially when combined. They allow us to test high-scale NP in motivated models that are still inaccessible via direct searches. Their experimental study should be pursued up to the few~\% level, i.e.~up to the level of the irreducible SM uncertainties, since they provide an invaluable tool to search or constrain 
new dynamics.

\section*{Acknowledgments} 
This project has received funding from the Swiss National Science Foundation~(SNF) under contract~200020\_204428.

\appendix

\section{Inputs}\label{sect:appendix}
In this appendix, we provide a complete list of the observables and the experimental inputs included in the analysis and we discuss the parametrization of NP effects in terms of effective 
Wilson coefficients. We focus solely on the operators and coefficients relevant to our SMEFT framework. More general discussions can be found in the cited references. 

\subsection{Collider observables}
We consider High-$p_T$ Drell-Yan tails, using LHC Run-II di-tau and mono-tau data. The likelihood is constructed using {\tt HighPT} \cite{Allwicher:2022mcg}, and running effects for these observables are neglected.

\subsection{Electroweak observables}
The Electroweak observables we consider are all the traditional $Z$- and $W$-pole observables as listed e.g. in \cite{Breso-Pla:2021qoe}. We use the same $\alpha$, $G_F$, and $m_Z$ input scheme and neglect SM uncertainties in all observables. New Physics contributions are computed in SMEFT in the linear approximation, i.e. only the dimension-six-SM interference terms are considered. For the explicit expressions of the observables in terms of SMEFT coefficients see \cite{Allwicher:2023aql}. 
Renormalization Group (RG) 
effects are taken into account in the first leading log approximation from a reference scale $\Lambda = 1$ TeV to $m_Z$.

\subsection{LFU in $\tau$ decays}
We consider lepton flavor universality in leptonic $\tau$ decays. The relevant observables are the ratios $\left|g_\tau / g_{\mu(e)}\right|$, defined as
\begin{equation}
\left|\frac{g_\tau}{g_{\mu(e)}}\right|  =\frac{\mathcal{B}(\tau \rightarrow e(\mu) \nu \bar{\nu}) / \mathcal{B}(\tau \rightarrow e(\mu) \nu \bar{\nu})_{\mathrm{SM}}}{\mathcal{B}(\mu \rightarrow e \nu\bar{\nu}) / \mathcal{B}(\mu \rightarrow e \nu\bar{\nu})_{\mathrm{SM}}}\,.
\end{equation}
The corresponding experimental values are reported by HFLAV \cite{HFLAV:2022esi}:
\begin{equation*}
\left|\frac{g_\tau}{g_\mu}\right|^{\textrm{exp}}  = (0.9 \pm 1.4)\,10^{-3} \,,\quad \left|\frac{g_\tau}{g_e}\right|^{\textrm{exp}}  = (2.7 \pm 1.4)\,10^{-3} \,,
\end{equation*}
with a correlation $\rho_{\tau}=0.51$.
NP effects can modify these ratios through the effective four leptons operators
\begin{equation}
\begin{split}
\mathcal{L}_{\textrm{eff}}\supset\, & C_{\nu \mu}(\bar{\nu}_L^{\mu}\gamma^{\alpha}\nu_L^{\tau})(\bar{\tau}_L\gamma_{\alpha}\mu_L)\\
+& C_{\nu e}(\bar{\nu}_L^{e}\gamma^{\alpha}\nu_L^{\tau})(\bar{\tau}_L\gamma_{\alpha}e_L)\,.
\end{split}
\end{equation}
Neglecting quadratic corrections, we can write with good accuracy 
\begin{equation}
    \left|\frac{g_\tau}{g_{\mu(e)}}\right|\simeq 1 - \frac{v^2}{2} \textrm{Re}[C_{\nu e(\mu)}]\,.
\end{equation}
The operators in Eq.~(\ref{eq:ops}) can contribute to these observables through RG mixing effects. Taking into account the leading-log evolution in the SMEFT:
\begin{equation}
C_{\nu \mu(e)}\simeq\frac{m_t^2 N_c}{8 \pi^2}\textrm{log}\left(\frac{\Lambda^2}{m_t^2}\right)C_{\ell q}^{3}(\Lambda)\,,
\end{equation}
where $\Lambda=1$ TeV is the UV scale , and $C_{\ell q}^{3}=(C_{\ell q}^{+}-C_{\ell q}^{-})/2$.

\subsection{LFU in $R_{D^{(*)}}$}
Semileptonic charged-current $b\to c \ell \nu$ transitions can be tested through the LFU ratios
$$
R_{D^{(*)}}=\frac{\mathcal{B}\left(B \rightarrow D^{(*)} \tau \nu\right)}{\mathcal{B}\left(B \rightarrow D^{(*)} \ell \nu\right)}\,.
$$
We use the HFLAV values \cite{HFLAV:2022esi} for both the experimental averages and the SM predictions of $R_{D^{(*)}}$, namely:
\begin{equation}
\begin{split}
&R_{D}^{\textrm{SM}}=0.298\pm0.004\,, \quad R_{D^{*}}^{\textrm{SM}}=0.254\pm0.005 \ , \\
&R_{D}^{\textrm{exp}}=0.342\pm0.026\, , \quad R_{D^{*}}^{\textrm{exp}}=0.287\pm0.012~,\\
& \hspace{2.9cm} \rho = -0.39~.
\end{split}
\end{equation}
Following Ref.\cite{Iguro:2022yzr,Duan:2024ayo}, we define the effective Lagrangian
\begin{equation}
    \begin{split}
        \mathcal{L}_{\rm eff}= - \frac{4 G_F V_{cb}}{\sqrt{2}}\big[&(1+C_{V_L})(\bar{c}\gamma^{\mu}P_Lb)(\bar{\tau}\gamma_{\mu}P_L\nu_{\tau})+\\
        &+C_{S_R}(\bar{c}P_Rb)(\bar{\tau}P_L\nu_{\tau}) \big]\,,
    \end{split}
\end{equation}
and simply parametrize the EFT dependence of these observables, normalized to the SM prediction, as:
\begin{equation}
    \begin{split}
        \frac{R_D}{R_D^{\rm{SM}}}&=|1+C_{V_L}|^2+1.01|C_{S_R}|^2+1.49\, \textrm{Re}[(1+C_{V_L})C_{S_R}^*]\,,\\
        \frac{R_{D^*}}{R_{D^*}^{\textrm{SM}}}&=|1+C_{V_L}|^2+0.04|C_{S_R}|^2-0.11\, \textrm{Re}[(1+C_{V_L})C_{S_R}^*]\,.
    \end{split}
\end{equation}
These low energy coefficients can be easily matched onto our SMEFT basis as follows:
\begin{equation}
\begin{split}
    C_{V_L}=&\frac{v^2}{ V_{cb}}(V_{cd}V_{td}^{*}\,\kappa\,\varepsilon\, C_{\ell q}^{3}+V_{cs}V_{ts}^*\,\varepsilon\, C_{\ell q}^{3}-V_{cb}  C_{\ell q}^{3})\,,\\
     C_{S_R}=&\frac{v^2}{2  V_{cb}}(V_{cd}V_{td}^{*}\,\kappa\,\varepsilon\, C_{S}+V_{cs}V_{ts}^*\,\varepsilon\, C_{S}-V_{cb}  C_{S})^*\,.
    \end{split}
    \label{eq:CVL}
\end{equation}
where $C_{\ell q}^{3}=(C_{\ell q}^{+}-C_{\ell q}^{-})/2$. The CKM factors $V_{ci}$ stem from the choice of working with a down-aligned basis and the $V_{ti}^*$ come from the insertion of spurions. The complex conjugation appears because, according to our definitions, $C_{S}$ multiplies the hermitian conjugate of the operator associated with $C_{S_R}$.

RG effects are significant in presence of scalar operators and thus they can not be neglected for this observable. We perform the running numerically using DSixTools \cite{Fuentes-Martin:2020zaz, Celis:2017hod}, evolving the operators up to the TeV scale.

\subsection{FCNC processes $b \to s\ell \bar\ell$}

The effective Lagrangian describing  $b\to s\ell\bar\ell$ ($\ell=e,\mu$) transitions, after integrating 
out the SM degrees of freedom above the 
$b$-quark mass, can be written as 
    \begin{equation}
    \label{eq:L_bsll}
       \cL_\mathrm{eff}(b\to s\ell\bar\ell) = 
        \frac{4 G_F}{\sqrt{2}}  V_{tb}V^*_{ts}
  \sum_{i=1}^{10} C_i Q_i \,,
    \end{equation}
    where 
    \begin{align}
        Q_9=&\frac{e^2}{16\pi^2}(\bar{s}_{L}\gamma_\mu b_L)(\bar{\ell}\gamma^\mu \ell)\,,  
    \end{align}
and the explicit form of all the other operators can be found in~\cite{Aebischer:2017gaw}.
Within our approach, NP induces vanishing tree-level contributions to all the $C_i$ in  (\ref{eq:L_bsll}). 
However, a sizable RG-induced contribution occurs from 
the QED running of the $(\bar b_L\gamma^\mu s_L) (\bar \tau_L \gamma_\mu \tau_L)$  
operator into  $Q_9$~\cite{Aebischer:2017gaw,Crivellin:2018yvo}. 
This leads to~\cite{Aebischer:2022oqe} 
\begin{equation}
\Delta C_9 \doteq 
 C^{\rm NP}_9  
\approx  \frac {\varepsilon  v^2 }{3}   C_{\ell q}^{+} 
\log{\left(
    \frac{\Lambda^2}{m^2_\tau}\right)}\,,
\end{equation}
to be compared with $C_9^{\rm SM} = 4.2$.
Based on the recent analyses of $B\to K^{(*)}\mu\bar\mu$ data in~\cite{Alguero:2022wkd,Bordone:2024hui}, taking into the SM uncertainties associated to charm-rescattering~\cite{Isidori:2024lng}, 
we set  
$\Delta C^{\rm exp}_9 =
C_9^{\rm exp}- C_9^{\rm SM} = - 0.6 \pm 0.2$.

\subsection{FCNC processes $d_i \to d_j \nu\bar\nu$}
We consider the neutrino golden-channel decays
\begin{equation*}
\begin{split}
\mathcal{R}^{BK^{(*)}}_{\nu\bar\nu}&=\frac{\mathcal{B}\left(B \rightarrow K^{(\ast)} \nu \bar{\nu}\right)}{\mathcal{B}\left(B
\rightarrow K^{(\ast)} \nu \bar{\nu}\right)_{\mathrm{SM}}}\,,\\[2mm]
\mathcal{R}^{K\pi}_{\nu\bar\nu}&=\frac{\mathcal{B}\left(K^{+} \rightarrow \pi^{+} \nu \bar{\nu}\right)}{\mathcal{B}\left(K^{+} \rightarrow \pi^{+} \nu \bar{\nu}\right)_{\mathrm{SM}}}\,,
\end{split}
\end{equation*}
mediated by the flavor-changing neutral current (FCNC) processes $b\to s \nu\bar\nu$ and $s\to d \nu\bar\nu$. 

For $\mathcal{R}^{BK}_{\nu\bar\nu}$ we use the experimental average that combines the recent Belle-II result with previous searches \cite{Belle-II:2023esi} and divide it by the SM prediction of \cite{Becirevic:2023aov}, resulting in $(\mathcal{R}^{BK}_{\nu\bar\nu})_{\rm exp} = 2.93\pm 0.92$. We also consider the upper limit $\mathcal{B}(B\to K^{*}\nu\bar{\nu})<2.7 \cdot 10^{-5}$ at the 90\% CL, as reported by Belle \cite{Belle:2017oht}. Assuming the SM expectation as central value, it yields $(R_{\nu\bar\nu}^{B K^*})_{\textrm{exp}}=1.0\pm1.1$.

Using the effective Lagrangian in Eq.~(\ref{eq:Leff}) and defining $\mathcal{R}^{BK^{(*)}}_{\nu\bar\nu}=1+\delta \mathcal{R}^{BK^{(*)}}_{\nu\bar\nu}$,  the NP contribution reads
\begin{equation}
    \delta \mathcal{R}^{BK^{(*)}}_{\nu\bar\nu} =\frac{2\mathrm{Re}[C_{\tau,bs}^{\mathrm{SM}}  C_{\tau,bs}]}{3 |C_{\tau,bs}^{\mathrm{SM}}|^2}+\frac{| C_{\tau,bs}|^{2}}{3|C_{\tau,bs}^{\mathrm{SM}}|^2}\,,
\end{equation}
where $C_{\tau,bs}^\textrm{SM}=-6.32(7)$. The tree level matching of the effective coefficient $C_{\tau,bs}$ onto the SMEFT basis is
$$C_{\tau,bs}=-\varepsilon\frac{\pi v^2}{ \alpha}C_{\ell q}^{-}\,.$$

For $\mathcal{R}^{K\pi}_{\nu\bar\nu}$ we use the SM prediction of Eq.~(\ref{eq:Ktopi}) where the numerical values of the coefficients are:
\begin{align*}
        P_c=&\,\left(\frac{0.2255}{\lambda}\right)^4 \times (0.3604\pm0.0087)\,,  \\
        \Delta_{\textrm{EM}}=&\,-0.003\,,   \qquad\qquad   \delta P_{c,u}=\,0.04(2)\,.
\end{align*}
The hadronic matrix elements and the coefficient $\kappa_+$ are discussed in detail in \cite{Mescia:2007kn,Bijnens:2007xa}. We use this estimate for $\mathcal{R}_{\nu \bar{\nu}}^{K \pi}$ along with the experimental average provided in \cite{NA62talk}, which includes the recent NA62 measurement. The ratio of these two values yields $(\mathcal{R}^{K\pi}_{\nu\bar\nu})_{\rm exp} = 1.6\pm0.4$. 
Again, calling $\mathcal{R}_{\nu \bar{\nu}}^{K \pi}=1+\delta \mathcal{R}_{\nu \bar{\nu}}^{K \pi}$, we can write the NP contribution as:
$$
    \delta \mathcal{R}_{\nu \bar{\nu}}^{K \pi} =\frac{2\mathrm{Re}[C_{\tau,sd}^{\mathrm{SM}}  C_{\tau,sd}]}{3 |C_{\tau,sd}^{\mathrm{SM}}|^2}+\frac{| C_{\tau,sd}|^{2}}{3|C_{\tau,sd}^{\mathrm{SM}}|^2}\,,
$$
where $C^{\rm SM}_{\tau,sd} =  C^{\rm SM}_{\tau,bs} - \frac{ \lambda^c_{sd} }{s_W^2 \lambda^t_{sd} } X^{\tau}_c$~\cite{Buchalla:1998ba,Brod:2008ss,Buras:2006gb,Brod:2010hi}. The matching of this low-energy coefficient to the SMEFT reads
$$C_{\tau,sd}=\varepsilon^2 \kappa \frac{\pi v^2}{ \alpha}C_{\ell q}^{-}\,.$$

\bibliography{references}
\end{document}